\documentclass[12pt]{article}

\usepackage{cite}
\begin{document}
\newcommand{\de}{\delta}
\newcommand{\ga}{\gamma}
\newcommand{\e}{\epsilon}
\newcommand{\ot}{\otimes}
\newcommand{\be}{\begin{equation}}
\newcommand{\ee}{\end{equation}}
\newcommand{\ba}{\begin{array}}
\newcommand{\ea}{\end{array}}
\newcommand{\beq}{\begin{equation}}
\newcommand{\eeq}{\end{equation}}
\newcommand{\tmod}{{\cal T}}
\newcommand{\amod}{{\cal A}}
\newcommand{\bemod}{{\cal B}}
\newcommand{\cmod}{{\cal C}}
\newcommand{\dmod}{{\cal D}}
\newcommand{\hmod}{{\cal H}}
\newcommand{\s}{\scriptstyle}
\newcommand{\tr}{{\rm tr}}
\newcommand{\einsop}{{\bf 1}}
\def\oR{R^*}
\def\upa{\uparrow}
\def\R{\overline{R}}
\def\doa{\downarrow}
\def\dag{\dagger}
\def\ve{\epsilon}
\def\si{\sigma}
\def\ga{\gamma}
\newcommand{\reff}[1]{eq.~(\ref{#1})}


\def\ba{\begin{array}}
\def\ea{\end{array}}
\def\no{\nonumber}
\def\le{\langle}
\def\re{\rangle}
\def\lt{\left}
\def\rt{\right}
\def\dwn{\downarrow}
\def\up{\uparrow}
\def\dag{\dagger}
\def\nonum{\nonumber}

\title{{Integrable generalised spin ladder models based on the
SU$(1|3)$ and SU$(3|1)$ algebras}}

\author{\large Arlei Prestes Tonel$^{1}$, Angela
Foerster$^1$, Katrina Hibberd$^2$\thanks{Corresponding author,
email: keh@posta.unizar.es }$~$  and Jon Links$^3$}

\date{}
\maketitle

\begin{center}
{\em ${}^1$Instituto de F\'{\i}sica da UFRGS \\
Av. Bento Gon\c{c}alves 9500, Porto Alegre, RS - Brazil }

{$ \phantom{0000000000}$}

{\em ${}^2$Departamento de F\'{\i}sica Te\'{o}rica, Facultad de Ciencias\\
Universidad de Zaragoza, 50009 Zaragoza, Spain}

{$ \phantom{0000000000}$}

{\em ${}^3$Centre for Mathematical Physics, School of Physical Sciences\\
The University of Queensland, QLD, 4072,  Australia}
\end{center}

\begin{abstract}
We present two integrable spin ladder models which possess a
general free parameter besides the rung coupling $J$. The models
are exactly solvable by means of the Bethe ansatz method and we
present the Bethe ansatz equations.  We analyse the elementary
excitations of the models which reveal the existence of a gap for
both models that depends on the free parameter.
 \end{abstract}
%
 \begin{flushleft}
 \vspace{1cm}
 \end{flushleft}
\vfil\eject

Spin ladder systems continue to attract attention motivated by
experimental realisations in quasi-one-dimensional systems
\cite{dagotto}.  These materials display novel features  and
with the continued development of new systems, there has been an
impressive amount of progress in the theoretical understanding of
such systems.  However, a greater flexibility through the
introduction of tunable free parameters within the well
established mathematical frameworks would be of considerable
advantage and forms the main aim of the present work.

It has been shown that ladder systems are reasonably well
approximated by Heisenberg interactions, which involve bilinear
exchanges \cite{dagotto1}.  While these models are not exactly
solvable, several more general systems have been proposed in which
solvability is guaranteed through the use of an extension of the
symmetry algebra \cite{frahmrod,albeverio,batchelor1,frahm}.
There has also been the introduction of systems involving
interactions beyond nearest neighbour exchanges which demonstrate
remarkably interesting behaviour and also prove to be exactly
solvable.  For example, the general 2-leg spin ladder system with
biquadratic interactions \cite{kolezhuk,nersesyan}.  The physical
importance of these types of interactions has been addressed in
\cite{honda}.

Subsequently other generalised integrable spin ladders have been
proposed
\cite{batchelor,angi,frahmstahl,scalapino,us}.
As is well known, integrability facilitates the use of long
established techniques in order to determine the physical
properties of such systems.  However, in these cases, no
free parameters other than the rung coupling
are present due to the strict conditions of
integrability.

In a recent article \cite{arlei} an
integrable model containing an additional free parameter was presented as
a generalisation of the model presented by Wang \cite{wang}. In this
instance, the free parameter plays the role of an anisotropy variable
and it was shown that the critical value of the rung coupling
which defines a Pokrovsky-Talapov phase transition between a gapped
and gapless phase was dependent on this anisotropy.
In \cite{albertini}, we note the study of a
family of spin ladder Hamiltonians which also have free parameters,
although in this case the construction has a different mathematical
origin.  It is clear
that this is a topic that warrants further investigation, since the
availability of tunable parameters yields a richer phase structure.

In this article, we present two new integrable generalised spin
ladders, based on the $SU(1|3)$ and $SU(3|1)$ symmetries, containing
an extra parameter.  The free parameter arises in the models as a
special choice of the multiparametric versions \cite{multi}. The
models are integrable in the sense that they contain an infinite
number of conservation laws and can be derived from a solution of
the Yang-Baxter equation.  This property is also of physical
importance as it provides a means to improve our understanding of
general correlated systems (see for example \cite{saleur}).  We
present the Bethe ansatz solution from which the physical
properties of the systems may be obtained.

An important characteristic of ladder systems, both from a
theoretical and experimental point of view, is the quantum phase
transition between gapped and gapless phases.  The spin gap is
vital for superconductivity to occur under doping, whilst from a
mathematical perspective, the size of the gap is dependent on the
relative strength of the rung
interaction parameter.  We address this issue as we analyse the
ground state and first excited states of the models.
Interestingly, we are able to show that for both systems a gap
persists in the spectrum of the elementary excitations and indeed
the gap depends on the extra parameter.

We begin by introducing the first generalised spin ladder model,
for which the explicit global Hamiltonian is of the form \be
H^{\s{{(1)}}}=\sum_{j=1}^{N} \biggl[ \, h_{j,j+1} +
{{\frac{1}{2}}} J \left( \vec{\sigma_{j}}.\vec{\tau_{j}}-1 \right)
\, \biggr], \label{ha} \ee and the local Hamiltonians are given by
\begin{eqnarray}
&h_{j,j+1}&=\frac{1}{4}(1+\sigma_{j}^{z}\sigma_{j+1}^{z})(1+\tau_{j}^{z}\tau_{j+1}^{z}) \,
+
(\sigma_{j}^{+}\sigma_{j+1}^{-}+\sigma_{j}^{-}\sigma_{j+1}^{+})(\tau_{j}^{+}\tau_{j+1}^{-}+
\tau_{j}^{-}\tau_{j+1}^{+})\quad \nonumber \\ & & +
 \frac{1}{2}(1+\sigma_{j}^{z}\sigma_{j+1}^{z})(t^{-1}\,\tau_{j}^{+}\tau_{j+1}^{-}+
t\, \tau_{j}^{-}\tau_{j+1}^{+})+
 \frac{1}{2}(t^{-1}\, \sigma_{j}^{+}\sigma_{j+1}^{-}+
t \,
\sigma_{j}^{-}\sigma_{j+1}^{+})(1+\tau_{j}^{z}\tau_{j+1}^{z})\quad
\nonumber \\ &&
-\frac{1}{8}(1+\sigma_j^z)(1+\sigma_{j+1}^z)(1+\tau_j^z)(1+\tau_{j+1}^z).
\nonumber
\end{eqnarray}
The parameters $\vec{\sigma_{j}}$ and $\vec{\tau_{j}}$ represent
Pauli matrices acting on site $j$ of the upper and lower legs
respectively, $J$ is the strength of the rung coupling that can
take arbitrary real values and $t$ is a general independent
parameter. The number of rungs is denoted by $N$ and periodic
boundary conditions are assumed.

The integrability of this model is assured by the Quantum Inverse
Scattering Method \cite{qism} and by the fact that it can be
mapped to the Hamiltonian given in equation (\ref{ha1}) below.
This Hamiltonian can be derived from an $R-$matrix obeying the
Yang-Baxter algebra \cite{baxter} for $J = 0$, while for $J\neq
0$, the rung interactions take the form of a chemical potential
term.  We find that
\begin{equation}
{\hat{H}}^{{\s(1)}} = \sum_{j=1}^{N} \biggl[ {\hat h}_{j,j+1} - 2J
\, X^{00}_{j} \biggr], \label{ha1}
\end{equation}
where
\begin{eqnarray}
&{\hat h}_{j,j+1}&=\sum_{\alpha =0}^{3}X^{\alpha
\alpha}_{j}X^{\alpha \alpha}_{j+1}+
X^{2 0}_{j}X^{0 2}_{j+1}+X^{0 2}_{j}X^{2 0}_{j+1} +
X^{1 3}_{j}X^{3 1}_{j+1}+X^{3 1}_{j}X^{1 3}_{j+1}  \quad \nonumber \\ & & +
t\biggl(X^{1 0}_{j}X^{0 1}_{j+1}+
X^{1 2}_{j}X^{2 1}_{j+1}+
X^{0 3}_{j}X^{3 0}_{j+1}+X^{2 3}_{j}X^{3 2}_{j+1}
\biggr) \quad \nonumber \\ & & +
t^{-1}\biggl(X^{0 1}_{j}X^{1 0}_{j+1}+
X^{2 1}_{j}X^{1 2}_{j+1}+
X^{3 0}_{j}X^{0 3}_{j+1}+X^{3 2}_{j}X^{2 3}_{j+1}
\biggr) -2X_j^{0 0} X_{j+1}^{0 0}.
\nonumber
\end{eqnarray}
In the above, $ X^{\alpha \beta}_{j} =
|\alpha_{j}\re\le\beta_{j}|$ are the Hubbard operators with $
|\alpha_{j}\re$ being the orthogonalised eigenstates of the local
operator $(\vec{\sigma_{j}}{\bf .} \vec{\tau_{j}})$.

The $R$-matrix we use is a special case of a more general
multiparametric version.  (For a general construction of
multiparametric models, see \cite{multi}.)  For the purposes of
the present work, it is necessary to only retain one parameter.
The $R$-matrix is as follows,
\begin{equation}
{\footnotesize R(x)=\pmatrix{w&0&0&0&|& 0&0&0&0&|&
0&0&0&0&|&0&0&0&0&\cr
               0&t^{-1}b&0&0&|& c&0&0&0&|& 0&0&0&0&|& 0&0&0&0&\cr
               0&0&b&0&|& 0&0&0&0&|& c&0&0&0&|& 0&0&0&0&\cr
               0&0&0&t b&|& 0&0&0&0&|& 0&0&0&0&|& c&0&0&0&\cr
               -&-&-&-& & -&-&-&-& & -&-&-&-& & -&-&-&-&\cr
               0&c&0&0&|& t b&0&0&0&|& 0&0&0&0&|& 0&0&0&0&\cr
               0&0&0&0&|& 0&a&0&0&|& 0&0&0&0&|& 0&0&0&0&\cr
               0&0&0&0&|& 0&0&t b&0&|& 0&c&0&0&|& 0&0&0&0&\cr
               0&0&0&0&|& 0&0&0&b&|& 0&0&0&0&|& 0&c&0&0&\cr
               -&-&-&-& & -&-&-&-& & -&-&-&-& & -&-&-&-&\cr
               0&0&c&0&|& 0&0&0&0&|& b&0&0&0&|& 0&0&0&0&\cr
               0&0&0&0&|& 0&0&c&0&|& 0&t^{-1}b&0&0&|& 0&0&0&0&\cr
               0&0&0&0&|& 0&0&0&0&|& 0&0&a&0&|& 0&0&0&0&\cr
               0&0&0&0&|& 0&0&0&0&|& 0&0&0&t b&|& 0&0&c&0&\cr
               -&-&-&-& & -&-&-&-& & -&-&-&-& & -&-&-&-&\cr
               0&0&0&c&|& 0&0&0&0&|& 0&0&0&0&|& t^{-1}b&0&0&0&\cr
               0&0&0&0&|& 0&0&0&c&|& 0&0&0&0&|& 0&b&0&0&\cr
               0&0&0&0&|& 0&0&0&0&|& 0&0&0&c&|& 0&0&t^{-1}b&0&\cr
               0&0&0&0&|& 0&0&0&0&|& 0&0&0&0&|& 0&0&0&a&\cr} } \,,
\end{equation}
with
$$a=x+1,\, \, \, b=x,\, \, \,  c=1,\,\,\, \mbox{and} \,\,\,  w=1-x,$$
and obeys the Yang-Baxter algebra \cite{baxter},
\begin{eqnarray}
R_{12}(x-y)R_{13}(x)R_{23}(y)=R_{23}(y)R_{13}(x)R_{12}(x-y).
\label{ybe}
\end{eqnarray}
From this solution originates the Hamiltonian (\ref{ha1}) for
$J=0$ by the standard procedure \cite{qism},
$$ {\hat h}_{j,j+1}= P \frac{d}{dx}R(x)|_{x=0}, $$
where $P$ is the permutation operator.

The model is exactly solvable by the Bethe ansatz method
\cite{bethe} and the resulting Bethe ansatz equations (BAE) are
given by the expressions,
\begin{eqnarray}
\label{bae1} -(-1)^{M_1} t^{(N-2M_3)}\left(\frac{\lambda_{l}-i/2}
{\lambda_{l}+i/2}\right)^{N}&=&
\prod_{j=1}^{M_{2}}\frac{\lambda_{l}-\mu_{j}-i/2}
{\lambda_{l}-\mu_{j}+i/2} ,\nonumber \\
t^{(N-2M_3)}\prod_{j\neq l}^{M_{2}}\frac{\mu_{l}-\mu_{j}-i}
{\mu_{l}-\mu_{j}+i}&=&
\prod_{i=1}^{M_{1}}\frac{\mu_{l}-\lambda_{i}-i/2}
{\mu_{l}-\lambda_{i}+i/2} \prod_{k
=1}^{M_{3}}\frac{\mu_{l}-\nu_{k}-i/2} {\mu_{l}-\nu_{i}+i/2},
 \\
t^{(N-2M_1+2M_2)}\prod_{k \neq l}^{M_{3}}
\frac{\nu_{l}-\nu_{k}-i}{\nu_{l}-\nu_{k}+i} &=& \prod_{j
=1}^{M_{2}}\frac{\nu_{l}-\mu_{j}-i/2} {\nu_{l}-\mu_{j}+i/2}.
\nonumber
\end{eqnarray}
The corresponding energy eigenvalues of the Hamiltonian
(\ref{ha1}) are,
\begin{equation}
\label{e1} { E}
=\sum_{j=1}^{M_{1}}\biggl(\frac{1}{\lambda_{j}^{2}+1/4}+2J\biggr)
-\left(1+2J\right)N,
\end{equation}
where $\lambda_{j}$ are solutions of the BAE (\ref{bae1}).

From the Bethe ansatz solution, we can determine the behaviour of
the ground state and elementary excitations of the system. The
reference state becomes the ground state when the relation
$J>-1+\frac{1}{2}(t+t^{-1})$ is satisfied.  For $N$ sites, the
ground state energy is $E_0=-\left(1+2J\right)N$, which in terms
of the Bethe ansatz calculations,  corresponds to the reference
state characterised by $M_1=M_2=M_3=0$.

To describe an elementary excitation, we chose $M_2=M_3=0$ and
$M_1=1$ in the BAE which, from equation (\ref{e1}), yields an
energy expression of the form
\begin{equation}
{ E}_1=\frac{1}{\lambda^2+1/4}+2J-(1+2J)N,
\end{equation}
where $\lambda=\frac{i}{2}\lt(\frac{t+1}{t-1}\rt)$. It is apparent
that there is a gap of
\begin{equation}
\Delta=2(J+1-\frac{1}{2}(t+t^{-1})).
\end{equation}
In the limit $t=1$, this solution corresponds to $\lambda
\rightarrow \infty$ indicating that a gap of $\Delta = 2J $
persists.  We note that this is agrees with the suggested
numerical and experimental results of spin ladder systems
\cite{dagotto}.


We move on to introduce the second integrable spin ladder model
which also contains a free parameter.  The global Hamiltonian
reads
\begin{equation}
\label{h2} {\cal{H}}^{{\s(2)}}=\sum_{j=1}^{N} \biggl[k_{j,j+1}+
{{\frac{1}{2}}} J \left( \vec{\sigma_{j}}.\vec{\tau_{j}}-1
\right)\biggr], \label{ha2}
\end{equation}
where
\begin{eqnarray}
&k_{j,j+1}&=\frac{1}{4}(1+\sigma_{j}^{z}\sigma_{j+1}^{z})(1+\tau_{j}^{z}\tau_{j+1}^{z}) \,
+
(\sigma_{j}^{+}\sigma_{j+1}^{-}+\sigma_{j}^{-}\sigma_{j+1}^{+})(\tau_{j}^{+}\tau_{j+1}^{-}
+\tau_{j}^{-}\tau_{j+1}^{+})\quad \nonumber \\ & & +
 \frac{1}{2}(1+\sigma_{j}^{z}\sigma_{j+1}^{z})(t^{-1}\,\tau_{j}^{+}\tau_{j+1}^{-}+
t\,\tau_{j}^{-}\tau_{j+1}^{+})+
 \frac{1}{2}(t^{-1}\,\sigma_{j}^{+}\sigma_{j+1}^{-}+
t\,\sigma_{j}^{-}\sigma_{j+1}^{+})(1+\tau_{j}^{z}\tau_{j+1}^{z})\quad
\nonumber \\ &&
-\frac{1}{8}(1-\sigma_j^z)(1-\sigma_{j+1}^z)(1-\tau_j^z)(1-\tau_{j+1}^z).
\nonumber
\end{eqnarray}

The exact solvability of the above Hamiltonian, as for the
previous case, lies in the fact that it too can be mapped to a
Hamiltonian given below by equation (\ref{ha21}).   Once again
this Hamiltonian is derived from an $R$-matrix solution of the
Yang-Baxter algebra for $J=0$, while for $J \neq 0$ the rung
interactions take the form of a chemical potential term.  The
Hamiltonian has the form,
\begin{eqnarray}
{\hat {\cal{H}}}^{{\s(2)}}=\sum_{j=1}^{N}\biggl[{\hat k}_{j,j+1}
-2J X^{00}_{j}\biggr], \label{ha21}
\end{eqnarray}
where
\begin{eqnarray}
&{\hat k}_{j,j+1}&=\sum_{\alpha =0}^{3}X^{\alpha
\alpha}_{j}X^{\alpha \alpha}_{j+1}+ X^{2 0}_{j}X^{0 2}_{j+1}+X^{0
2}_{j}X^{2 0}_{j+1} + X^{1 3}_{j}X^{3 1}_{j+1}+X^{3 1}_{j}X^{1
3}_{j+1}  \quad \nonumber \\ & & + t\biggl(X^{1 0}_{j}X^{0
1}_{j+1}+ X^{1 2}_{j}X^{2 1}_{j+1}+ X^{0 3}_{j}X^{3 0}_{j+1}+X^{2
3}_{j}X^{3 2}_{j+1} \biggr) \quad \nonumber \\ & & +
t^{-1}\biggl(X^{0 1}_{j}X^{1 0}_{j+1}+ X^{2 1}_{j}X^{1 2}_{j+1}+
X^{3 0}_{j}X^{0 3}_{j+1}+X^{3 2}_{j}X^{2 3}_{j+1} \biggr) -2X_j^{3
3} X_{j+1}^{3 3}. \nonumber
\end{eqnarray}

For $J=0$, the model is derived, in a similar manner as for the
above case, from a multiparametric ${R}$-matrix for which only one
parameter is relevant for the present discussion. The $R$-matrix is
given by
\begin{equation}
{\footnotesize R(x)=\pmatrix{a&0&0&0&|& 0&0&0&0&|& 0&0&0&0&|&
0&0&0&0&\cr
               0&t^{-1}b&0&0&|& c&0&0&0&|& 0&0&0&0&|& 0&0&0&0&\cr
               0&0&b&0&|& 0&0&0&0&|& c&0&0&0&|& 0&0&0&0&\cr
               0&0&0&t b&|& 0&0&0&0&|& 0&0&0&0&|& c&0&0&0&\cr
               -&-&-&-& & -&-&-&-& & -&-&-&-& & -&-&-&-&\cr
               0&c&0&0&|& t b&0&0&0&|& 0&0&0&0&|& 0&0&0&0&\cr
               0&0&0&0&|& 0&a&0&0&|& 0&0&0&0&|& 0&0&0&0&\cr
               0&0&0&0&|& 0&0&t b&0&|& 0&c&0&0&|& 0&0&0&0&\cr
               0&0&0&0&|& 0&0&0&b&|& 0&0&0&0&|& 0&c&0&0&\cr
               -&-&-&-& & -&-&-&-& & -&-&-&-& & -&-&-&-&\cr
               0&0&c&0&|& 0&0&0&0&|& b&0&0&0&|& 0&0&0&0&\cr
               0&0&0&0&|& 0&0&c&0&|& 0&t^{-1}b&0&0&|& 0&0&0&0&\cr
               0&0&0&0&|& 0&0&0&0&|& 0&0&a&0&|& 0&0&0&0&\cr
               0&0&0&0&|& 0&0&0&0&|& 0&0&0&t b&|& 0&0&c&0&\cr
               -&-&-&-& & -&-&-&-& & -&-&-&-& & -&-&-&-&\cr
               0&0&0&c&|& 0&0&0&0&|& 0&0&0&0&|& t^{-1}b&0&0&0&\cr
               0&0&0&0&|& 0&0&0&c&|& 0&0&0&0&|& 0&b&0&0&\cr
               0&0&0&0&|& 0&0&0&0&|& 0&0&0&c&|& 0&0&t^{-1}b&0&\cr
               0&0&0&0&|& 0&0&0&0&|& 0&0&0&0&|& 0&0&0&w&\cr} } \,
,
\end{equation}
with
$$a=x+1,\, \, \, b=x,\, \, \,  c=1\, \, \,\mbox{and }    w=-x+1,$$
and satisfies the Yang-Baxter algebra (\ref{ybe}). Utilising  the
Bethe ansatz method this model can be solved and the resulting BAE
are,
\begin{eqnarray}
\label{bae2}
t^{(N-2M_3)}\left(\frac{\lambda_{l}-i/2}
{\lambda_{l}+i/2}\right)^{N}&=&  \prod_{l \neq i}^{M_{1}}\frac{\lambda_{l}-\lambda_{i}-i}
{\lambda_{l}-\lambda_{i}+i}
\prod_{j=1}^{M_{2}}\frac{\lambda_{l}-\mu_{j}+i/2}
{\lambda_{l}-\mu_{j}-i/2} ,\nonumber \\
t^{(N-2M_3)}\prod_{j\neq l}^{M_{2}}\frac{\mu_{l}-\mu_{j}-i}
{\mu_{l}-\mu_{j}+i}&=&
\prod_{i=1}^{M_{1}}\frac{\mu_{l}-\lambda_{i}-i/2}
{\mu_{l}-\lambda_{i}+i/2} \prod_{k
=1}^{M_{3}}\frac{\mu_{l}-\nu_{k}-i/2} {\mu_{l}-\nu_{i}+i/2},
 \\
-(-1)^{M_3}t^{-(N-2M_1+2M_2)} &=& \prod_{j
=1}^{M_{2}}\frac{\nu_{l}-\mu_{j}-i/2} {\nu_{l}-\mu_{j}+i/2}
..\nonumber
\end{eqnarray}
The eigenenergies of the Hamiltonian  (\ref{h2}) are given by
\begin{equation}
{\cal E}=-\sum_{j=1}^{M_{1}}\biggl
(\frac{1}{\lambda_{j}^{2}+1/4}-2J \biggr)+(1-2J)N,
\end{equation}
where $\lambda_{j}$ are solutions of BAE (\ref{bae2}).

For $N$ sites, the ground state is given by a product of  rung
singlets  when \\$J >1+\frac{1}{2}(t+t^{-1})$ and the energy is
${\cal E}_0=(1-2J)N$. This is in fact the reference state used in
the Bethe ansatz calculations and corresponds to the case
$M_1=M_2=M_3=0$ of the BAE (\ref{bae2}). To describe an elementary
spin-$1$ excitation, we choose $M_1=1 $ and $M_2=M_3=0$ in the BAE
which gives the minimal excited state energy,
\begin{equation}
{\cal E}_1=-\frac{1}{\lambda_{1}^{2}+1/4}+2J+\left(1-2J\right)N,
\end{equation}
where $\lambda_{1}=\frac{i}{2}\lt(\frac{t-1}{t+1}\rt)$. The energy
gap can easily be calculated and is found to be
\begin{equation}
\Delta_1=2\biggl(J-1-\frac{1}{2}(t+t^{-1})\biggr).
\end{equation}
The value $J^c=1+\frac{1}{2}(t+t^{-1})$ indicates the critical
line at which the transition from dimerized phase to the gapless
phase occurs.


In conclusion, we have presented two new spin ladder models
derived as special cases of multiparametric versions of $SU(3|1)$
and $SU(1|3)$ invariant solutions of the Yang-Baxter equation, maintaining
 one  free parameter besides the rung coupling $J$.
Upon investigation of the solutions of the BAE`s to determine
ground state and elementary excitations, we have shown that both
models exhibit a gap that depends on the extra parameter. Our results
show
similar generic properties to the $SU(4)$ model studied in \cite{arlei}
and is very suggestive that such multiparametic extensions will, in
general, always have an influence on the physical characteristics of
these models, and in particular the critical value of the rung coupling.
~~\\

\centerline{{\bf Acknowledgements}}
~~\\
AF and APT thank CNPq-Conselho Nacional de
Desenvolvimento Cient\'{\i}fico e Tecnol\'ogico for financial support.
KEH acknowledges the support of the Ministerio de Educaci\'on y
Cultura, Espa\~na. JL thanks the Australian Research Council.


\end{document}